\documentclass[aps,amssymb,showpacs]{revtex4}
\usepackage{amssymb}
\usepackage{graphicx}
\usepackage{color}
\usepackage{textcomp}
\usepackage{ulem}
\begin{document}

\title{Inverse Magnetic Catalysis in Nambu--Jona-Lasinio Model beyond Mean Field }
\author{Shijun Mao}
\affiliation{School of Science, Xian Jiaotong University, Xian 710049, China}

\begin{abstract}
We study inverse magnetic catalysis in the Nambu--Jona-Lasinio model beyond mean field approximation. The feed-down from mesons to quarks is embedded in an effective coupling constant at finite temperature and magnetic field. While the magnetic catalysis is still the dominant effect at low temperature, the meson dressed quark mass drops down with increasing magnetic field at high temperature due to the dimension reduction of the Goldstone mode in the Pauli-Villars regularization scheme.\\

\end{abstract}


\date{\today}
\maketitle

The Quantum Chromodynamics (QCD) phase transition in an external magnetic field has drawn much attention in recent years, due to its close relation to high energy nuclear collisions~\cite{hi0,hi1,hi2,hi3,hi31,hi32,hi4,hi5,hi6}, compact stars~\cite{ns0,ns1,ns2,ns21,ns3,ns4,ns41,ns42,ns43,ns5,ns6,ns61} and cosmological phase transitions~\cite{cpt1,cpt2,cpt3,cpt4}. Considering the dimension reduction of fermions, the chiral symmetry breaking is enhanced by the magnetic field, which leads to an increasing critical temperature for the chiral restoration phase~\cite{mc1,mc2,mc3}. However, the lattice simulation of QCD performed with physical pion mass observes the opposite phenomenon, namely the critical temperature drops down with increasing magnetic field~\cite{la1,la2,la3,la4}. Many scenarios are proposed to understand this inverse magnetic catalysis~\cite{fukushima,kamikado,bf1,bf11,bf12,bf13,bf2,bf3,bf4,bf5,bf6,bf7,bf8,bf9,bf10,bf101}, such as the magnetic inhibition of mesons, the mass gap in the large $N_c$ limit, the sphalerons, the gluon screening effect, and the weakening of the strong coupling.

In the NJL model without external magnetic field, the mean field approximation for quarks together with the random phase approximation for mesons can describe well the chiral thermodynamics of hot and dense quark-meson plasma~\cite{njl1,njl2,njl3,njl4,njl5}, and the feed-down from mesons to quarks leads to a lower critical temperature~\cite{zhuang}. With a strong magnetic field, the Goldstone mode in the chiral symmetry breaking phase may play an important role for the realization of inverse magnetic catalysis~\cite{fukushima}. One problem in the NJL model is the regularization. Since the model with contact interaction among quarks is nonrenormalizable, one requires a regularization scheme to avoid the divergent momentum integrations. When the external magnetic field is turned on, the quark energy becomes discrete and the phase space becomes anisotropy. In this case taking a proper regularization scheme becomes significant to guarantee the law of causality, see the following discussion.

In this Letter, we focus on the inverse magnetic catalysis resulted from the Goldstone mode at strong magnetic field. In the framework of the NJL model, we will calculate the effective quark coupling constant by including the feed-down from the Goldstone mode and solve the corresponding gap equation for the chiral condensate. We will see that the regularization scheme here plays an important role.

The $SU(2)$ NJL model is defined through the Lagrangian density~\cite{njl1,njl2,njl3,njl4,njl5}
\begin{equation}
{\cal L}=\bar{\psi}\left(i\gamma_{\nu} D^{\nu}-m_0\right)\psi + \frac{G}{2}\left[\left( \bar{\psi} \psi \right)^2 + \left( \bar{\psi} i \gamma_5 {\vec \tau} \psi \right)^2  \right], \label{gs}
\end{equation}
where the covariant derivative $D^{\nu}=\partial^\nu+i Q A^\nu$ couples quarks to the external magnetic field ${\bf B}=(0, 0, B)=\nabla\times{\bf A}$ along the $z$-axis, $Q=diag(Q_u, Q_d)=diag(\frac{2}{3}e,-\frac{1}{3}e)$ is the quark charge matrix in flavor space, and $G$ is the coupling constant in the scalar and pseudo-scalar channels. In chiral limit with vanishing current quark mass $m_0=0$, the $SU(2)_L \otimes SU(2)_R$ symmetry is broken down to $U(1)_L \otimes U(1)_R$ by the magnetic field ${\bf B}$, and the number of Goldstone modes is reduced from 3 to 1. In the chiral symmetry breaking phase, quarks obtain mass $m=m_0-G \langle \bar\psi\psi\rangle$ from the chiral condensate $\langle \bar\psi\psi \rangle$.

With the Leung-Ritus-Wang method~\cite{ritus1,ritus2,ritus21,ritus3,ritus5,ritus6,ritus7,ritus9}, the quark propagator with flavor $f$ in coordinate space can be written as
\begin{eqnarray}
S_f(x,y) &=& i\sum_{n=0}^\infty \int {d\tilde p\over (2\pi)^3} e^{-i \tilde p\cdot (x-y)} P_n(x_1,p_2)D_f(\bar p) P_n(y_1,p_2),\nonumber\\
P_n(z,q) &=& {1\over 2}\left[g_n^{s_f}(z,q)+I_n g_{n-1}^{s_f}(z,q)\right]+{is_f\over 2}\left[g_n^{s_f}(z,q)- I_n g_{n-1}^{s_f}(z,q)\right]\gamma^1 \gamma^2,\nonumber\\
D_f^{-1}(\bar p) &=& \gamma \cdot \bar p-m,
\label{ritusp}
\end{eqnarray}
where $\bar p=(p_0,0,-s_f \sqrt{2|Q_f B|n},p_3)$ is the Ritus momentum with the sign factor $s_f=\text{sgn}(Q_f B)$ for $f=u,d$, the magnetic field dependent function $g_n^{s_f}(z,q) = \phi_n(z-s_f q /|Q_f B|)$ is controlled by the Hermite polynomial $H_n(z)$ via $\phi_n(z) = \left(2^n n! \sqrt{\pi} |Q_f B|^{-1/2}\right)^{-1/2} e^{-z^2|Q_f B|/2} H_n\left(z/|Q_f B|^{-1/2}\right)$, and the Fourier transformed momentum $\tilde p$ and the Landau energy level factor $I_n$ are defined as $\tilde p=(p_0,0,p_2,p_3)$ and $I_n=1-\delta_{n0}$.

In mean field approximation, the thermodynamic potential of the system at finite temperature $T$, baryon chemical potential $\mu_B$ and external magnetic field $B$ includes the mean field part and the quark part,
\begin{eqnarray}
\Omega_{mf} &=& \frac{m^2}{2 G}+\Omega_q,\nonumber\\
\Omega_q &=& -3 \sum_{f=u,d}\sum_n \alpha_n \int \frac{d p_z}{2\pi} \frac{|Q_f B|}{2\pi}\left[\frac{E_f^++E_f^-}{2}+ T \ln \left(\left(1+e^{-E_f^+/T}\right)\left(1+e^{-E_f^-/T}\right)\right)\right]
\end{eqnarray}
with the spin factor $\alpha_n=2-\delta_{n0}$ and quark energies $ E^\pm_f=\sqrt{p^2_z+2 n |Q_f B|+m^2}\pm \mu_B/3$. Note that with the replacement $\sum_ n |Q_f B|/(2\pi)\alpha_n \int d p_z/(2\pi) \longrightarrow \ 2\int d^3 {\bf p}/(2\pi)^3$, we can recover the thermodynamic potential without magnetic field~\cite{zhuang}. The physical quark mass or the chiral condensate is determined by minimizing the thermodynamic potential $\partial\Omega_{mf}/\partial m=0$ which leads to the gap equation in chiral limit,
\begin{equation}
m\left( \frac{1}{2G}+\frac{\partial \Omega_q}{\partial m^2}\right)=0.
\label{gapeq}
\end{equation}

Now we go beyond the mean field approximation by considering the meson contribution to the thermodynamic potential.
In NJL model, mesons are treated as quantum fluctuations and constructed through random phase approximation~\cite{njl1,njl2,njl3,njl4,njl5}. There are four kinds of mesons, the isospin singlet $\sigma$ and triplet $\pi_0$ and $\pi_\pm$, via interactions in the scalar and pseudoscalar channels. The meson polarization function, namely the quark bubble, is defined as
\begin{equation}
\Pi_{M}(k) = -i \int d^4 (x-x') e^{i k \cdot (x-x')}{\text {Tr}}\left[\Gamma_M S(x,x') \Gamma_M^* S(x',x)\right]
\end{equation}
with the meson vertex
\begin{equation}
\label{vertex} \Gamma_M = \left\{\begin{array}{ll}
1 & M=\sigma\\
i\tau_+\gamma_5 & M=\pi_+ \\
i\tau_-\gamma_5 & M=\pi_- \\
i\tau_3\gamma_5 & M=\pi_0\ ,
\end{array}\right.\ \
\Gamma_M^* = \left\{\begin{array}{ll}
1 & M=\sigma\\
i\tau_-\gamma_5 & M=\pi_+ \\
i\tau_+\gamma_5 & M=\pi_- \\
i\tau_3\gamma_5 & M=\pi_0\
\end{array}\right.
\end{equation}
and the quark propagator matrix in flavor space $S=diag(S_u,\ S_d)$, where the trace is done in spin, color and flavor spaces. Via taking the quark bubble summation in random phase approximation, the meson propagator can be written as
\begin{equation}
D_M(k)=\frac{G}{1-G\Pi_M(k)},
\end{equation}
and the meson pole mass $m_M$ and the quark-meson coupling constant $g_{q\bar q M}$ are defined at the pole of the propagator at zero momentum~\cite{njl1,njl2,njl3,njl4,njl5,zhuang,ritus5,ritus6},
\begin{eqnarray}
\label{pole}
&& 1-G\Pi_M(k_0^2=m^2_M,{\bf k}^2=0)=0,\nonumber\\
&& \left(g^\mu_{q\bar q M}\right)^2 =\left[g^{\mu\mu} \frac{d \Pi_M (k)}{d k^2_\mu}\bigg|_{k^2=(m^2_M, 0)}\right]^{-1}
\end{eqnarray}
with $g^{\mu\nu} = diag(1,-1,-1,-1)$.

After a straightforward calculation, the polarization function can be expressed as
\begin{equation}
\Pi_M(k)= 12i \sum_{n,l=0}^\infty \int \frac{dq_0 dq_2 d q_3}{(2\pi)^3} d x_1 e^{-i k_1 x_1}\Lambda^M_{nl}(x_1,q,k).
\end{equation}
For neutral mesons $M=\sigma, \pi_0$, the integrated function $\Lambda_{nl}^M$ becomes relatively simple,
\begin{eqnarray}
\Lambda^M_{nl}(x_1,q,k) &=& \sum_{f=u,d}\frac{1}{(\bar p^2-m^2)(\bar q^2-m^2)}\bigg[2 \bar p_2 \bar q_2\beta_{nlf}^-(x_1,p_2)\beta_{nlf}^-(0,q_2)\nonumber\\
&&+ \left(\bar q \cdot \bar p+ \kappa_M m^2\right)\left(\beta_{nlf}^+(x_1,p_2)\beta_{nlf}^+(0,q_2)+ \beta_{nlf}^-(x_1,p_2) \beta_{nlf}^-(0,q_2)\right)\bigg]
\end{eqnarray}
with the definitions $p_j=k_j+q_j$ for $j=0,2,3$, $\kappa_{\pi_0}=-1$, $\kappa_{\sigma}=1$, and
\begin{equation}
\beta_{nlf}^\pm(z,q) = \frac{1}{2} \left[g^{s_f}_n(z,q)g^{s_f}_l(z,q) \pm I_n I_l g^{s_f}_{n-1}(z,q)g^{s_f}_{l-1}(z,q)\right].
\end{equation}
Taking into account the orthonormal relations for $g^{s_f}_n$,
\begin{eqnarray}
&& \int dz g^{s_f}_n(z,p)g^{s_f}_l(z,q)|_{p=q}=\delta_{nl}, \nonumber \\
&& \int dp g^{s_f}_n(0,p)g^{s_f}_l(0,q)|_{p=q}=|Q_f B|\delta_{nl},
\end{eqnarray}
the polarization function $\Pi_M$ for $M=\sigma, \pi_0$ at the pole can be simplified as
\begin{equation}
\Pi_M(k_0^2,0) = 3\sum_{f=u,d} \sum_{n=0}^\infty \alpha_n \left|\frac{Q_f B}{2\pi}\right| \int \frac{d p_z}{2\pi}\frac{E_f^2-\epsilon_M^2/4}{E_f^2-k_0^2/4} \frac{\tanh(\frac{E_f^+}{2T})+\tanh(\frac{E_f^-}{2T})}{E_f^+ + E_f^-}
\label{nmeson}
\end{equation}
with $\epsilon_{\pi_0}=0$ and $\epsilon_\sigma=2m$. By comparing the gap equation (\ref{gapeq}) for mean field quark mass $m_{mf}$ with the pole equation (\ref{pole}) for neutral meson masses $m_M$, we have the simple relations in the chiral symmetry breaking phase,
\begin{equation}
m_{\pi_0} = 0,\ \ \ \ \ m_\sigma = 2m_{mf}.
\end{equation}
This indicates that $\pi_0$ is the Goldstone mode corresponding to the spontaneous chiral symmetry breaking. Note that the magnetic field reduces the $SO(1,3)$ Lorentz group of space-time transformation to its $SO(1,1)$ subgroup in the direction of the field, which leads to an anisotropic coupling constant $g^\mu_{q {\bar q} M}$ with elements $g^1_{q {\bar q} M}=g^2_{q {\bar q} M}\neq g^0_{q {\bar q} M}=g^3_{q {\bar q} M}$~\cite{ritus5,ritus6}.

Including meson degrees of freedom in the model, the thermodynamic potential of the quark-meson plasma can be generally written as
\begin{equation}
\Omega={m^2\over 2G}+\Omega_q+\sum_M\Omega_M.
\end{equation}
Under pole approximation, mesons are quasi-particles and their thermodynamical potential can be simply expressed as
\begin{equation}
\Omega_M = \int \frac{d^3 {\bf k}}{(2\pi)^3} \left[\frac{E_M}{2} +T \ln\left(1-e^{-E_M/T}\right)\right]
\label{omeson}
\end{equation}
with meson energy
\begin{equation}
E_M = \sqrt{m_M^2+k_3^2+v^2_\perp (k_1^2+k_2^2)},
\end{equation}
where the meson mass $m_M$ is determined by the pole equation (\ref{pole}), $v^2_\perp =\left(g^0_{q {\bar q} M}\right)^2/\left(g^1_{q {\bar q} M}\right)^2$ is called transverse velocity~\cite{ritus5,ritus6}, and $v^2_\perp \neq 1$ indicates the anisotropy in meson energy dispersion relations due to the introduction of the external magnetic field.

The physical quark mass as the order parameter of chiral phase transition corresponds to the minimum of the thermodynamic potential of the system at fixed temperature, chemical potential and magnetic field, $\partial\Omega(m,T,\mu_B,B)/\partial m=0$. In mean field approximation, this leads to the gap equation (\ref{gapeq}) and determines the mean field quark mass $m_{mf}$. Going beyond the mean field, the feed-down from mesons to quarks results in a new term in the gap equation,
\begin{equation}
m\left( \frac{1}{2G}+\frac{\partial \Omega_q}{\partial m^2}+\sum_M\frac{\partial \Omega_M}{\partial m^2}\right)=0.
\label{gapnew}
\end{equation}
Obviously, the new order parameter $m$ from the new gap equation (\ref{gapnew}) is different from the mean field one $m_{mf}$, and the difference comes from the quantum fluctuations above the mean field.

Suppose the fluctuations induced correction is small, $|m-m_{mf}|/m_{mf}<<1$, we can expand the meson thermodynamics in terms of the correction~\cite{zhuang},
\begin{equation}
\Omega_M=\sum_n{\frac{1}{n!} \frac{\partial^n \Omega_M}{\partial (m^2)^n}\Big|_{m_{mf}^2}\left(m^2-m_{mf}^2\right)^n}.
\end{equation}
To simplify the calculation, we keep only the first two terms of this series with $n=0,1$. Under this approximation, the gap equation takes the same form as the mean field one, and the meson correction is reflected in an effective coupling constant $G'$,
\begin{eqnarray}
&& m\left( \frac{1}{2G'}+\frac{\partial \Omega_q}{\partial m^2}\right) =0,\nonumber\\
&& \frac{1}{2G'} =\frac{1}{2G}+\sum_M\frac{\partial \Omega_M}{\partial m^2}\Big|_{m^2_{mf}}.
\label{gnew}
\end{eqnarray}
Different from the original coupling $G$ which is a constant, the effective coupling $G'$ is defined in the medium and the external field. It is its dependence on both the magnetic field and the temperature that leads to the magnetic catalysis at low temperature and inverse magnetic catalysis around the critical temperature, see Fig.\ref{fig3} in the following. This is in line with the conclusion~\cite{g3} that chiral models having couplings with a magnetic field dependence only fail to describe the inverse magnetic catalysis.

Because of the four-fermion interaction, NJL model is not a renormalizable theory and needs regularization. Different from directly introducing a hard or soft three-momentum cutoff to the quark momentum which is widely used in cases without~\cite{njl2,njl3,njl4,njl5} and with~\cite{soft} magnetic field, the Pauli-Villars regularization scheme is covariant and describes well the quark and meson masses~\cite{florkowski} and quark potential~\cite{mu}. In this scheme, the quark momentum runs formally from zero to infinity, and the divergence is removed by the cancellation among the subtraction terms. While the magnetic field does not cause extra divergence, it introduces discrete Landau level and anisotropy in momentum space. In this case, the Pauli-Villars scheme can guarantee the law of causality. Under the Pauli-Villars scheme, one introduces the regularized quark masses $m_i=\sqrt{m^2+a_i\Lambda^2}$ for $i=0,1,\cdots, N$, and replaces $m^2$ in the quark energy $E_f$ by $m_i^2$ and the summation and integration $\sum_n\int dp_z/(2\pi)F(E_f)$ by $\sum_n\int dp_z/(2\pi)\sum_{i=0}^N c_iF(E_f^i)$. The coefficients $a_i$ and $c_i$ are determined by constraints $a_0=0$, $c_0=1$, and $\sum_{i=0}^N c_im_i^{2L}=0$ for $L=0,1,\cdots N-1$. This treatment to cancel the divergence is very different from the procedure in the soft-cutoff scheme~\cite{soft} where the integrated function $F(E_f)$ is simply multiplied by a cutoff factor $U_\Lambda=\Lambda^{2N}/\left[\Lambda^{2N}+(p_z^2+2n |Q_f B|)^N\right]$.

There are two parameters $G$ and $\Lambda$ in the NJL model in chiral limit, which are determined by fitting the pion decay constant $f_\pi=93$ MeV and chiral condensate $\langle \bar \psi \psi \rangle =(-250 \text{MeV})^3$ in vacuum. The obtained parameters are $N=5$, $G=14.36$ GeV$^{-2}$ and $\Lambda=638.8$ MeV in soft-cutoff scheme and $N=3$, $G=9.94$ GeV$^{-2}$ and $\Lambda=1127$ MeV in Pauli-Villars scheme. Note that the fitting is through the new gap equation (\ref{gnew}) instead of the mean field one~\cite{zhuang}.

To calculate the phase transition line of chiral symmetry restoration at finite temperature and magnetic field, we focus in this Letter on the chiral breaking phase where the meson degrees of freedom play the dominant role. Since charged pions $\pi_\pm$ and $\sigma$ are massive at nonzero magnetic field, their contribution to the thermodynamics of the system in chiral breaking phase is much smaller in comparison with the Goldstone mode $\pi_0$, especially when the magnetic field is strong. To avoid the complicated calculation for charged pions, we consider only quarks and the Goldstone mode $\pi_0$. To guarantee this approximation being good enough, we take in the following the restrict on the magnetic field $20<eB/m_\pi^2<50$ where $m_\pi$ is the pion mass in vacuum. In this case, the charged pions and $\sigma$ are heavy and the application of the NJL model is still reasonable satisfying the constraint $eB/m_\pi^2 < \Lambda^2$ in the Pauli-Villars regularization.
\begin{figure}[hb]
\centering
\includegraphics[width=7.5cm]{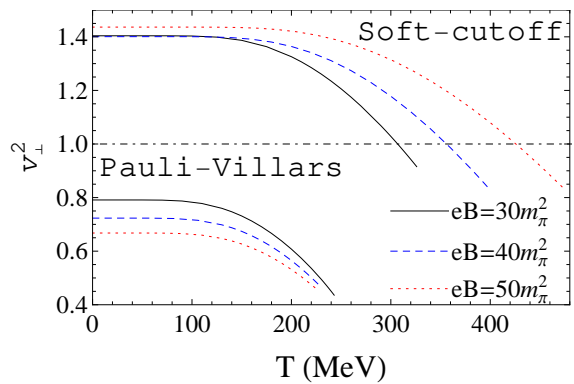}
\caption{The transverse velocity $v_\perp^2$ for the Goldstone mode $\pi_0$ as a function of temperature at different magnetic field in the chiral breaking phase. The Pauli-Villars (lower panel) and Soft-cutoff (upper panel) regularization schemes are used.}
\label{fig1}
\end{figure}

We first solve the gap equation (\ref{gapeq}) to obtain the mean field quark mass $m_{mf}$ and then substitute it into the pole equation (\ref{pole}) to calculate the quark-meson coupling constant $g^{\mu}_{q\bar q M}$ and transverse velocity $v_\perp$. Note that there is $m_{\pi_0}=0$ in the whole chiral breaking phase. For the Goldstone mode $\pi_0$, its longitudinal velocity is exactly the speed of light, $v_{||}=1$, and its transverse velocity $v_{\perp}$ is shown in Fig.\ref{fig1} at finite temperature and magnetic field. In Pauli-Villars regularization scheme, $v_\perp$ is always less than the speed of light, satisfying the law of causality. With increasing magnetic field, the transverse motion becomes more and more slow. The result of $v_\perp <1$ for massless particles means a dimension reduction and is expected to happen in external magnetic field. Applying the Mermin-Wagner-Coleman theorem~\cite{coleman1,coleman2,coleman3} which forbids the spontaneous breaking of continuous symmetries in space with dimension less than two, the dimension reduction for the Goldstone mode leads to the possibility of inverse magnetic catalysis in quark-meson plasma. We will prove this numerically in Fig.\ref{fig4}. With the soft-cutoff scheme, however, the Goldstone mode shows some strange properties. While there exists still anisotropy, there is no more dimension reduction even at extremely strong magnetic field. The transverse velocity is larger than the speed of light at low temperature, violating the law of causality, which is the case even for very weak magnetic field~\cite{ritus5,ritus6}.

\begin{figure}[hb]
\centering
\includegraphics[width=7.5cm]{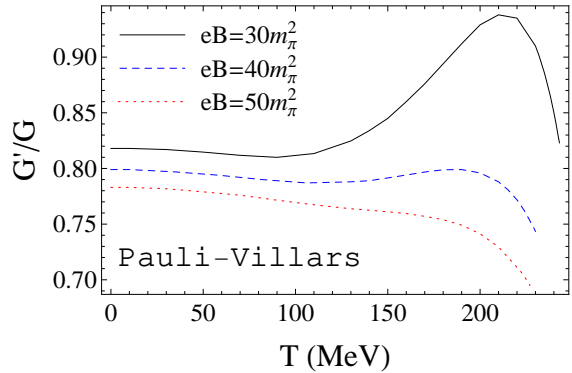}
\caption{The scaled effective coupling constant $G'/G$ as a function of temperature at different magnetic field in the chiral breaking phase. The Pauli-Villars regularization scheme is used. }
\label{fig2}
\end{figure}

\begin{figure}[hb]
\centering
\includegraphics[width=7.5cm]{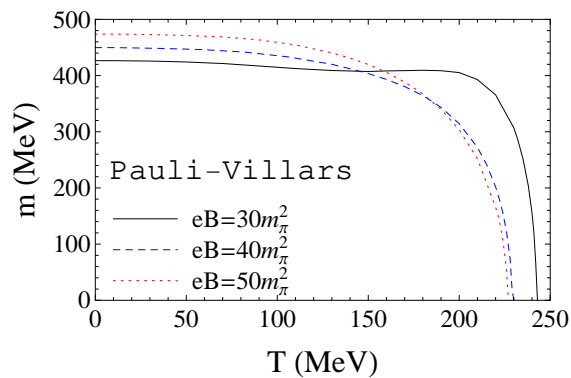}
\caption{The quark mass $m$ beyond mean field as a function of temperature at different magnetic field in the chiral breaking phase. The Pauli-Villars regularization scheme is used. }
\label{fig3}
\end{figure}

\begin{figure}[hbt]
\centering
\includegraphics[width=7.5cm]{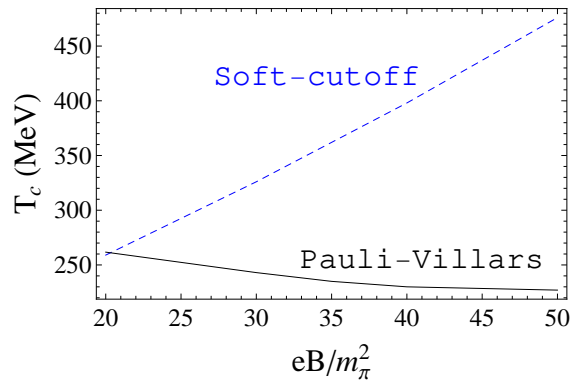}
\caption{The critical temperature $T_c$ of chiral symmetry restoration beyond mean field as a function of magnetic field in Pauli-Villars (solid lines) and soft-cutoff (dashed lines) regularization schemes. }
\label{fig4}
\end{figure}

With the mean field quark mass $m_{mf}$ and the meson pole masses $m_M$, we calculate the effective coupling constant $G'$ through the definition (\ref{gnew}) which is essential to control the magnetic field effect on chiral phase transition. Fig.\ref{fig2} shows the scaled effective coupling $G'/G$ at finite temperature and magnetic field in the Pauli-Villars regularization scheme. The feed-down effect of the mesons weakens the coupling among quarks with $G'/G<1$. At fixed temperature, the coupling drops down with increasing magnetic field, indicating the magnetic inhabition effect of mesons discussed in Ref.~\cite{fukushima}.

With the known effective coupling $G'$, we now recalculate the quark mass beyond mean field approximation. When the feed-down from mesons to quarks is fully embedded in the quark coupling, the quark-meson system is treated as an effective quark system. From the new gap equation (\ref{gnew}), the magnetic field effect on the quark mass shows two opposite aspects: The field in the quark thermodynamic potential $\Omega_q$ plays the role of magnetic catalysis through the dimension reduction of quarks, which enhances the quark mass, and the field in the effective coupling $G'$ shows magnetic inhibition through the dimension reduction of mesons, which reduces the quark mass due to the weakened attractive interaction $G'<G$. These two aspects are both temperature dependent and their competition controls the behavior of the quark mass. Fig.\ref{fig3} shows the quark mass at finite temperature and magnetic field. At low temperature, the magnetic catalysis is dominant, and the quark mass $m$ goes up with increasing magnetic field. At high temperature, however, the magnetic inhibition plays the dominant role, and the quark mass drops down with increasing magnetic field and becomes saturated when the field is strong enough.

The critical temperature $T_c$ of chiral symmetry restoration is defined by the definition $m(T_c, \mu_B, B)=0$. While the low temperature behavior of the meson dressed quark mass is still governed by the magnetic catalysis, it is controlled by the magnetic inhibition around the critical temperature, see Fig.\ref{fig3}. Therefore, we expect an inverse magnetic catalysis effect on the critical temperature $T_c$. The calculated magnetic field dependence of $T_c$ in Pauli-Villars regularization scheme is shown in Fig.\ref{fig4}. It drops down monotonously with increasing magnetic field in the region $eB<50 m_\pi^2$, and the behavior is in agreement with the lattice QCD simulations~\cite{la1,la2,la3,la4}. At mean field level, the critical temperature at vanishing magnetic field is about 300 MeV in the Pauli-Villars regularization scheme~\cite{florkowski,mu}. With different parameter values or choosing other regularization schemes, the critical temperature can be reduced. For instance, in the three-momentum non-covariant scheme, the critical value is $T_c$ = 170 MeV~\cite{njl2,njl3,njl4,njl5}. As a comparison, we show also in Fig.\ref{fig4} the critical temperature in the soft-cutoff regularization scheme, see the dashed line. Contrary to the Pauli-Villars scheme, $T_c$ goes up with increasing magnetic field due to the breaking of the law of causality shown in Fig.\ref{fig1}. It is clear that the dimension reduction of mesons plays an important role in the inverse magnetic catalysis phenomena.

In this Letter we investigated the magnetic field effect on the phase transition of chiral symmetry restoration in the frame of $SU(2)$ NJL model with different regularization schemes. We go beyond the mean field approximation by introducing an effective coupling which includes the feed-down from mesons to quarks, especially from the Goldstone mode $\pi_0$ at strong magnetic field. In Pauli-Villars regularization scheme, we observed the dimension reduction of $\pi_0$ which leads to a decreasing effective coupling with increasing magnetic field. As a consequence of the competition between the magnetic catalysis at mean field level and magnetic inhibition for the Goldstone mode, the order parameter of the phase transition is enhanced at low temperature but suppressed around the critical temperature by the magnetic field. Our result agrees qualitatively with the lattice QCD simulations.

From the comparison with the soft-cutoff regularization, the Pauli-Villars regularization plays an important role in obtaining the inverse magnetic catalysis. To see if the covariance controls the calculation, one needs to compare with other regularization schemes, such as the magnetic field independent regularization (MFIR)~\cite{mfir1,mfir3} which successfully separates the magnetic contribution from the vacuum thermodynamic potential and especially avoids the unphysical oscillations at high density. We should also point out that the meson correction to the quark self-energy considered here is only in the random phase approximation. It can be considered as a $1/N_c$ correction, but it is not a complete set. There are other well-developed $1/N_c$ expansion schemes in the NJL model, such as the scheme with a separable non-local interaction~\cite{blaschke}, the scheme through iterating the quark self-energy at Hartree approximation~\cite{oertel}, the scheme with saddle-point expansion~\cite{nikolov}, and the scheme with $\Phi$-derivable theory with nonlocal contributions~\cite{muller}. Since including the meson degrees of freedom reduces the critical temperature of chiral phase transition~\cite{zhuang,blaschke,muller}, we expect an inverse magnetic catalysis in these 1/Nc expansion schemes. Finally, we need to further calculate the observable quantities such as the magnetic field dependence of the pion mass, pion decay constant, and make comparison with lattice simulation~\cite{lattice} and MFIR regularizationl~\cite{mfir4}. The calculation at finite baryon density is especially important, since it is closely related to the magnetic field effect in compact stars and lattice simulation fails to do it at the moment.

\noindent {\bf Acknowledgement:} The work is supported by the NSFC and China Postdoctoral Science Foundation Grants 11405122, 11575093 and 2014M550483.

\end{document}